\documentclass[12pt,preprint]{aastex}
\def\kms{{\rm km\,s^{-1}}}

\def\masyr{{\rm mas}\,{\rm yr}^{-1}}

\def\rpm{{\rm RPM}}
\def\lim{{\rm lim}}

\begin{document}

\title{Transit Target Selection Using Reduced Proper Motions}

\author{Andrew Gould and Christopher W.\ Morgan}
\affil{Department of Astronomy, The Ohio State University,
140 W.\ 18th Ave., Columbus, OH 43210}
\email{gould,morgan@astronomy.ohio-state.edu}

\singlespace

\begin{abstract}

In searches for planetary transits in the field, well over half of the
survey stars are typically giants or other stars that are too large
to permit straightforward detection of planets.  For all-sky searches
of bright $V\la 11$ stars, the fraction is $\sim 90\%$.  We show that the great
majority of these contaminants can be removed from the sample by
analyzing their reduced proper motions (RPMs): giants have much lower
RPMs than dwarfs of the same color.  We use Hipparcos data to design
a RPM selection function that eliminates most evolved stars, while
rejecting only 9\% of viable transit targets.  Our method can be applied
using existing or soon-to-be-released all-sky data to stars $V<12.5$ in the
northern hemisphere and $V<12$ in the south.  The method degrades at fainter
magnitudes, but does so gracefully.  For example, at $V=14$ it can still
be used to eliminate giants redward of $V-I\sim 0.95$, that is,
the blue edge of the red giant clump.

\end{abstract}
\clearpage
 
\section{Introduction
\label{sec:intro}}

	Spurred on by the detection of the transiting planetary companion 
to HD209458 \citep{char00}
as well as the exciting scientific results that can be extracted
from intensive followup of this object \citep{char02,cody02,hui02},
a large number of surveys are underway to detect planetary transits
of stars both in clusters \citep{str00}
and the field \citep{how00,brown99,mal01,udal02,str02}.
 The expected signature of a planetary transit,
a $\sim 1\%$ drop in the stellar flux over a duration of several hours,
can be mimicked by a variety of non-planetary phenomena.  These
include transits by brown dwarfs or late M dwarfs, which have sizes
similar to Jovian planets, grazing eclipses by ordinary stars,
full eclipses by binaries that are $\sim 100$ times fainter than
the target star but lie in the same spatial resolution element, and
transits of evolved stars by main-sequence stars.  While it is sometimes
possible to distinguish grazing eclipses from the flatter-bottomed
transits using lightcurves of sufficient photometric precision,
rejection of stellar eclipses and transits usually requires
several spectra: an M star companion induces radial velocity (RV)
variations of $\sim 10\,\kms$, whereas planetary perturbations are
10 to 100 times smaller.

	Since obtaining these spectra is expensive in both telescope
time and human effort, it is important to seek other robust methods
of recognizing non-planetary transits.  Here we show how the
great majority of evolved stars can be eliminated from the target list of
transit searches using reduced proper motion (RPM) diagrams.  Field star 
transit surveys usually contain one to several times more evolved stars
than main-sequence stars, while only the latter present useful targets.
Hence, robust rejection of evolved-star contamination of transit
surveys should result in substantial gains in efficiency in both
the data analysis and the follow-up observations required for confirmation.

	In \S~\ref{sec:principles}, we review the basic physical principles
that allow one to recognize evolved stars from a RPM diagram.  
In \S~\ref{sec:tycho}, we assemble an
an all-sky catalog of transit targets $V_T<11$.  We show that our RPM
criteria reject about 60\% of the stars in this magnitude range that 
remain after an initial 25\% are already rejected because their colors
are too blue.  That is, altogether 70\% are rejected.
In \S~\ref{sec:2mass}, we establish the criteria for rejecting evolved
stars using a combination of Tycho-2 \citep{t2} and 2MASS \citep{2mass}
data.  These criteria
can be applied to stars $V\la 12$ over the whole sky once the full 2MASS 
catalog is released.
In \S~\ref{sec:ucac}, we show how the same technique can be extended
to $V\sim 12.5$ over the northern sky by combining 2MASS and UCAC \citep{ucac}
data, supplemented with data from the transit surveys themselves.  We also
show that the method degrades gracefully at fainter magnitudes, and can 
still be used to eliminate the majority of evolved stars at $V=14$.
While most of the paper focuses on the problem of selecting targets for
transits of amplitude $\sim 1\%$, we consider the possibility in
\S~\ref{sec:bigger} that smaller amplitudes might be reached for brighter
subsamples of the main survey.  In this way, efficient
transit searches might be extendable to stars of somewhat larger radius.
Finally, in \S~\ref{sec:discuss}, we briefly discuss our results.

\section{Principles
\label{sec:principles}}

	Most ground-based transit surveys aim to detect transits with
depths of about 1\%, roughly the fraction of solar light blocked by
Jupiter.  Since planets, or even brown dwarfs, are not expected to
get much larger than Jupiter, this level of sensitivity limits the
parent-star population that can be probed with transits to those with 
radii $r\la r_\odot$.  Early main-sequence (MS) stars
and stars that have evolved significantly off the MS are not useful:
1\% transits would then be due to other stars, not planets.  

	Broadly speaking, the contaminants can be divided into three
spectral groups: early-type, mid-type (roughly ``G-type''), 
and late-type.  It is straightforward 
to eliminate early-type stars using a simple color cut, and in
fact most survey teams already do this.  Late-type stars come in two distinct
classes, giants and dwarfs, which are separated by several magnitudes
in absolute magntitude.  In general, the parallaxes (and hence absolute
magnitudes) of these stars are unknown, but if their proper motions are
measured with sufficient accuracy, these classes can nevertheless
be distinguished using a reduced proper motion (RPM) diagram.  
The $V$ band RPM, $V_\rpm$, is related to $M_V$ by 
\begin{equation}
V_\rpm \equiv V + 5\log\mu = M_V + 5\log{v_\perp\over 47.4\,\kms},
\label{eqn:vrpm}
\end{equation}
where $\mu$ is the proper motion in $\rm arcsec\,\rm yr^{-1}$ and
$v_\perp$ is the transverse velocity.  Hence, a RPM diagram looks similar
to a standard color-magnitude diagram (CMD) but with greater scatter
owing to the dispersion in $v_\perp$.
However, for colors at which the giants and dwarfs are separated by
$\ga 3\,$mag, this scatter is not sufficient to confuse the two populations.

	Contamination from evolved G stars poses the most difficult
problem.  The MS and evolved populations are typically separated by just
a couple of magnitudes.  This is enough to make the evolved stars
too large to permit transit detections, 
but not large enough to easily distinguish them
from dwarfs of the same color using a RPM diagram.  Some contamination
from evolved G stars is therefore inevitable.

\begin{figure}
%\epsscale{0.7}
\plotone{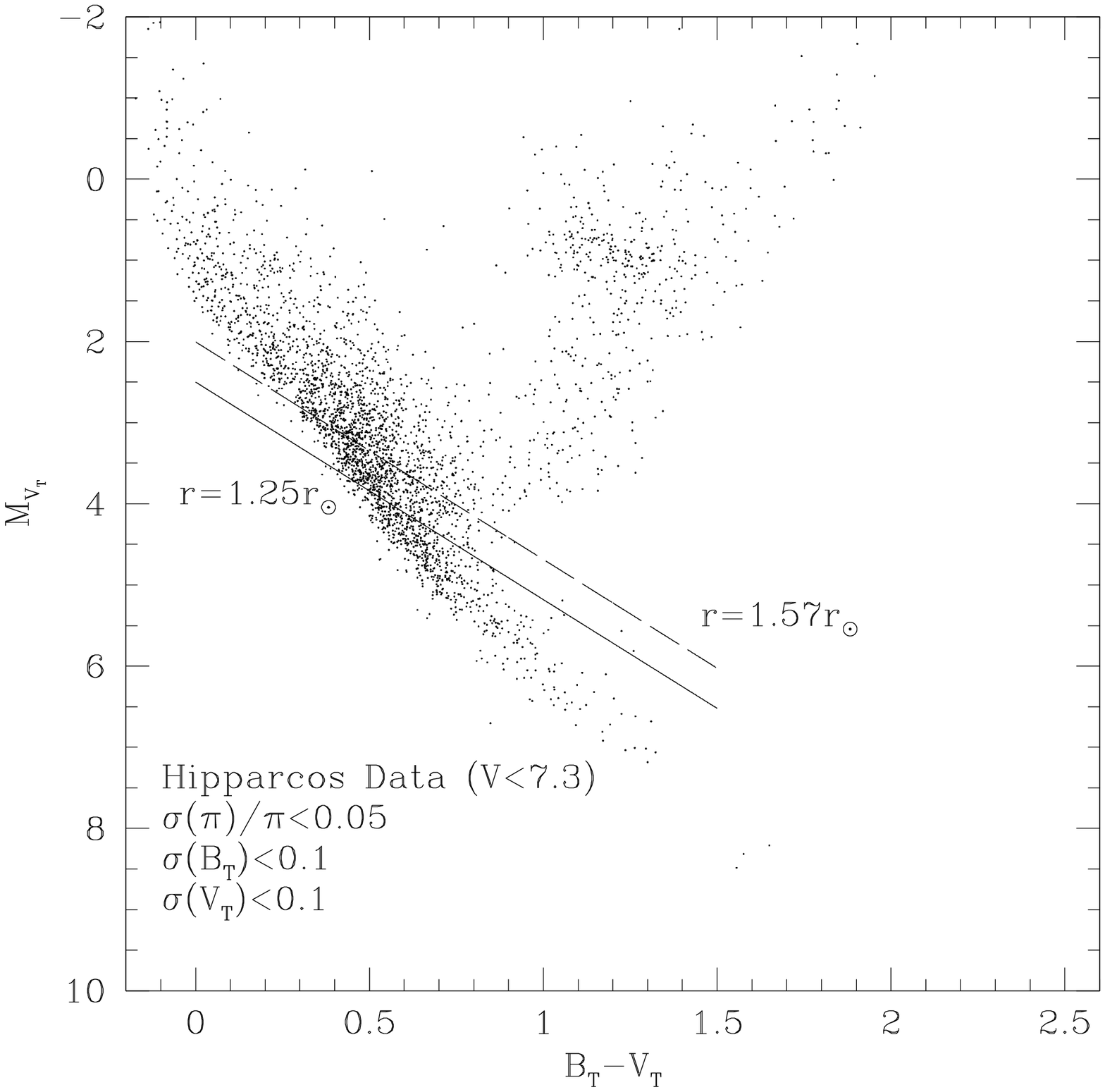}
\caption{\label{fig:hipcmd}
Hipparcos color-magnitude diagram (CMD)
restricted to stars $V<7.3$ with parallax errors smaller than 5\%
and photometry errors smaller than 0.1 mag.  The solid and dashed lines are
contours of constant radius, respectively $r=1.25\,r_\odot$ and 
$r=1.57\,r_\odot$.  Our primary objective is to design criteria that
efficiently select stars below the solid line, but without making use
of parallax information.
}\end{figure}

	Figure \ref{fig:hipcmd} is a CMD of Hipparcos stars
within the completeness limit $V<7.3$ and further restricted to parallax 
errors below 5\% and photometry errors below 0.1 mag in each band.  
The solid line shows the location of stars with estimated
radii $r=1.25\,r_\odot$, for which we use the formula,
\begin{equation}
\log {r\over r_\odot} = 0.597 + 0.536(B_T - V_T) - {M_{V_T}\over 5}.
\label{eqn:rstar}
\end{equation}
Here, $B_T$ and $V_T$ are the Hipparcos passbands.  To find the slope
in this equation, we combine the \citet{vanb} scaling,
$\log r \propto 0.064(V-K) - 0.2\,M_K$, with the standard Hipparcos \citep{hip}
transformation $(V-V_T)=-0.09(B_T-V_T)$, and a transformation
$(V_T-K) = 2.0525(B_T-V_T) + 0.1588$ empirically determined by matching
Hipparcos and 2MASS \citep{2mass} data over the relevant range 
$0.4<(B_T-V_T)<0.8$.  We normalize the zero point to the radius of the
Sun, assuming it has $(B_T-V_T)_\odot=0.71$ and $M_{V_T,\odot}=4.89$, which we
obtain from the \citet{hip} transformations starting with 
$(B-V)_\odot=0.65$, $M_{V,\odot}=4.83$.  

	In Figure \ref{fig:hiprpm}, we plot the RPMs of all the stars
from Figure \ref{fig:hipcmd}, the upper and lower panels showing
respectively the stars with large $(r>1.25\,r_\odot)$ and small
$(r<1.25\,r_\odot)$ radii.  The object now is to design color/RPM criteria
that will select the overwhelming majority of small stars while rejecting
as many large stars as possible.  The broken lines in the two panels
show our attempt to define such a selection function.  

	What properties must a dataset have to allow one to apply this  
selelction
function to it?  The most important criterion is that stars with 
$r\sim 1.25\,r_\odot$ (i.e., $M_V\sim 4$)
must have proper motion errors that are substantially smaller than their
typical proper motions.  If their proper motions were measured to be
consistent with zero,
it would not be possible to distinguish them from evolved stars of the
same color and apparent magnitude, which would then also have measured
proper motions
consistent with zero.  Disk stars have typical transverse velocities
$v_\perp\sim 40\,\kms$.  Therefore stars with $M_V=4$ have typical proper 
motions,
\begin{equation}
\mu \sim 21\,\masyr \times 10^{0.2(12-V)}.
\label{eqn:mv4lim}
\end{equation}
Second, the colors must be accurate enough to minimize scatter across
the color boundary at $(B_T-V_T) = 0.45$.

\section{Application to the Tycho-2 Catalog
\label{sec:tycho}}

An obvious choice for such a dataset is the Tycho-2 catalog \citep{t2}.
At $V=12$, Tycho-2 has typical proper motion errors of 
$\sigma_\mu\sim 3.5\,\masyr$, which
according to equation (\ref{eqn:mv4lim}) is quite adequate.  Tycho-2 is
also complete to $V\sim 12$.  However, Tycho-2 photometry,
particularly $B_T$ photometry is quite poor at these faint magnitudes.
In our initial treatment, we will therefore restrict consideration to
Tycho-2 stars with $V_T<11$, for which $\sigma_\mu\la 2.5\,\masyr$.
To avoid clutter, we display in Figure \ref{fig:tycrpm} a random 2\% 
subset of these stars.  Table 1 shows the total number of stars in the sample
that are accepted and rejected, with percentages broken down into
early-type $(B_T-V_T<0.45)$, mid-type $(0.45<B_T-V_T<0.80)$, and late-type
 $(B_T-V_T>0.80)$ stars.  If we ignore the 
stars that are rejected solely because they were too blue $(B_T-V_T<0.45)$
(on the grounds that such color selection was already being applied prior
to our work), we find that 60\% of the remaining stars are rejected by our
selection criteria.  

\begin{figure}
%\epsscale{0.7}
\plotone{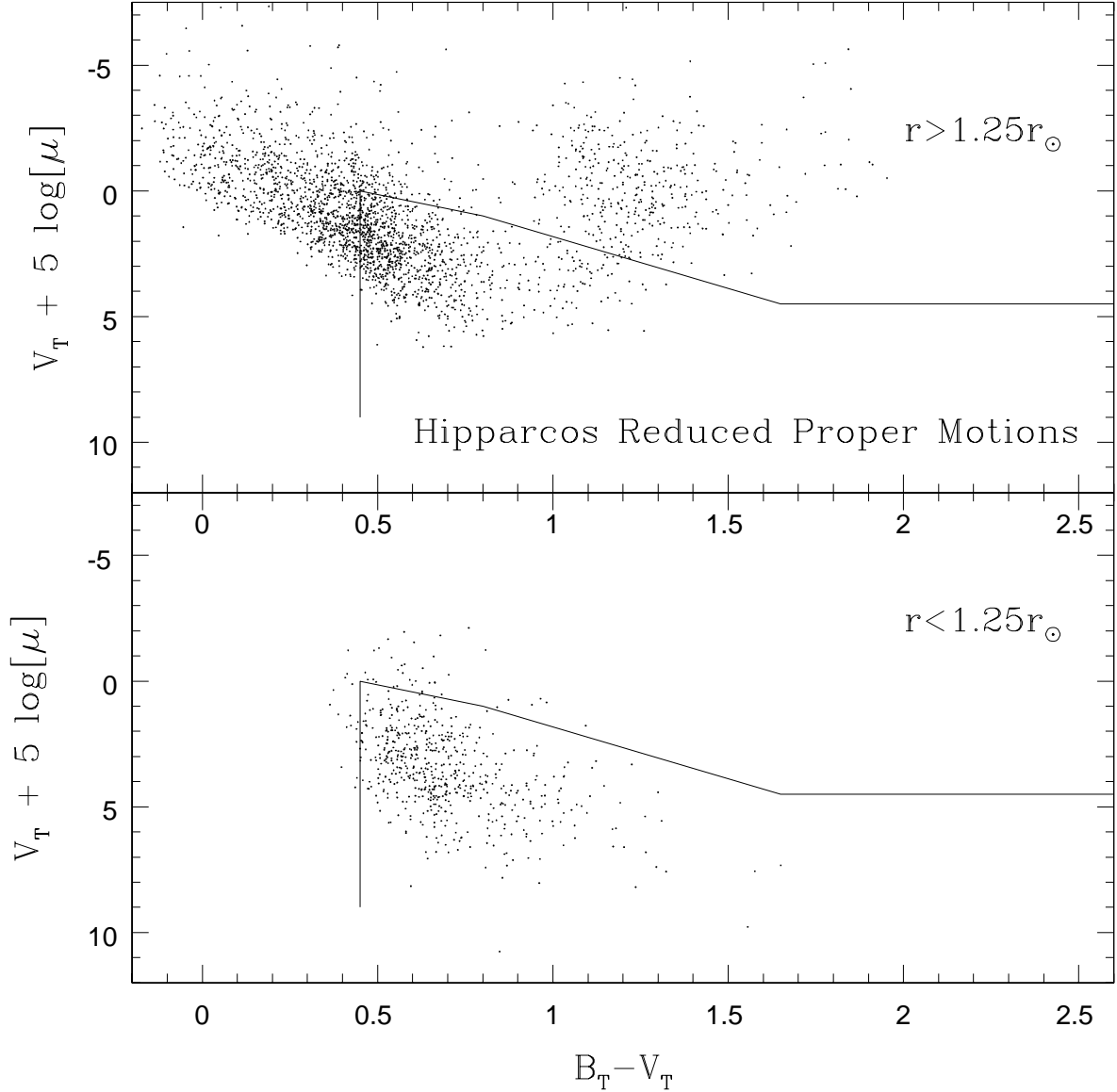}
\caption{\label{fig:hiprpm}
Hipparcos reduced proper motion (RPM) diagram.  
The upper and lower panels show respectively
the stars above and below the solid ($r=1.25\,r_\odot$) line in 
Fig.~\ref{fig:hipcmd}.  The broken lines (same in both panels) show our
adopted selection criteria for distinguishing stars $r<1.25\,r_\odot$.
The false rejection rate is about 9\%.  The contamination rate appears
to be about 65\%, but is somewhat underestimated because the underlying
sample is biased against intrinsically bright stars ($M_{V_T}\ga 3.3$) 
relative to a magnitude-limited sample.
}\end{figure}

	Of course, a successful selection will not only reject many
unwanted stars, but also keep the overwhelming majority of desirable stars.
It will also minimize the number of unwanted stars that are accepted.
We must use the stars in
Figure \ref{fig:hiprpm} to evaluate performance in these two areas,
since it is only this sample for which we have independent knowledge of
whether an individual star is desirable.  This will require some care
because Figures \ref{fig:hiprpm} and \ref{fig:tycrpm} are selected in 
somewhat analogous, but not identical, ways.

The fraction of stars falsely rejected in Figure \ref{fig:hiprpm} is
only 9\% (50/[536 +50]).  This is likely to be a pretty accurate estimate
of the false rejection rate for the Tycho-2 sample.  Since all the
stars in question have $M_V\ga 4$, the Hipparcos sample lies within 
$\sim 50\,$pc. Since Hipparcos parallax errors at these bright magnitudes
are $\la 1\,$mas, the fractional errors are $\la 5\%$.  Hence the
5\% parallax-error criterion does not substantially affect the sample
of stars $r<1.25\,r_\odot$.  Thus, both the Hipparcos and Tycho-2 samples 
are effectively 
magnitude-limited.  Because the Tycho-2 sample has a fainter magnitude
limit, it has a slightly lower fraction of stars near the maximum
allowed radius, since these stars are seen to a distance $\sim 250\,$pc
from the Galactic plane where their density thins out somewhat.  Hence,
the 9\% false rejection rate is probably a slight overestimate.

	The rate of unwanted
acceptances is 65\% (985/[985+536]).  This is likely to be somewhat
underestimated because the unwanted stars are intrinsically brighter
than the desired stars and so have suffered more from the bias induced by
demanding good parallaxes.  This bias is probably not extremely severe because
to be accepted, the unwanted star must be reasonably close to the boundary.

	Note that, since 70\% of stars are rejected and only $\la 35\%$ of 
those accepted are useful, the fraction of stars in the overall 
magnitude-limited sample that are useful is only 10\%.  

	Table 1 also shows the results for a sample limited to $V_T<10$, 
but otherwise selected according to the same criteria. This would be similar
to the sample probed by the all-sky $1''$ telescope survey proposed by
\citet{pgd}.  In addition, Table 1 shows
a $V_T<10$ sample selected according to looser criteria as described in
\S~\ref{sec:bigger}.

\section{Application to Tycho-2 + 2MASS
\label{sec:2mass}}

	The results in \S~\ref{sec:tycho} were restricted to $V_T<11$.
Since most transit surveys go beyond this limit, we would like
to push the method to fainter stars.  Recall from equation
(\ref{eqn:mv4lim}) that at $V=12$, the Tycho-2 astrometry errors 
are adequate ($3.5\,\masyr$) but the $B_T$ photometry is unreliable.
A straightforward way to resolve this problem is to match the Tycho-2
catalog to 2MASS and so replace $(B_T-V_T)$ by $(V_T -J)$.  At these
magnitudes, the 2MASS $J$ errors are typically $<0.03\,$mag and so
negligible compared to Tycho-2 errors.  Also, although the $V_T$ errors
do deteriorate significantly for these faint stars, the effects of
this are mitigated by the longer wavelength baseline: 
$d(V_T -J)/d(B_T-V_T)\sim 1.73$.  Unfortunately, the full 2MASS catalog
is not yet available, but it is promised by the end of 2002.  In the meantime,
the Second Incremental 2MASS Release, which covers 47\% of the sky, allows
us to evaluate the efficiency of this approach.

	Figure~\ref{fig:2massrpm} is a Tycho-2/2MASS RPM diagram, with 
$V_{T,\rpm}$ determined from Tycho-2, and colors from Tycho-2 $V_T$
and 2MASS $J$.  A random $4.25\%(=2\%/47\%)$ of the data are shown so that
the density of points can be compared directly to that of
Figure~\ref{fig:tycrpm}.  The
broken line discriminates between transit target stars and non-target
stars.  The end points of these line segments are found by
transforming the coordinates of the line segments in Figures~\ref{fig:hiprpm}
and \ref{fig:tycrpm}.  To determine this transformation, we fit $(V_T-J)$
of Hipparcos/2MASS stars to a fifth-order polynomial of $(B_T-V_T)$.
This fit has coefficients
$(0.064677, 1.915935,-0.034922,-0.676674, 0.457714,-0.075894)$
with a residual scatter of 0.113 mag.
No magnitude limit is applied to Figure~\ref{fig:2massrpm}.  
(Note that the numbers of accepted and rejected stars have been scaled
up by a factor 2.13 in Table 1 to take account of the fact that 2MASS
presently covers only 47\% of the sky.)

\section{Fainter Surveys
\label{sec:ucac}}

	With the advent of UCAC2 \citep{ucac}, it should be possible to
push our method to about $V=12.5$, at least in the north 
($\delta>-2.\hskip-2pt ^\circ 5$).  For these declinations, UCAC2 expects
to be able to link up modern CCD astrometry with archival AGK2 plates
$(B<13)$ to achieve $1\,\masyr$ proper motions (N.\ Zacharias 2002, private
communiation). For the most critical colors, $(B-V)\sim 0.5$, this
corresponds to $V=12.5$.  Such high astrometric
precision would easily satisfy the
requirements of equation (\ref{eqn:mv4lim}).  Unfortunately, UCAC2
does not expect their CCD photometry to be calibrated, and there is no
other source of all-sky optical CCD photometry at these magnitudes.
However, a transit survey could calibrate its own relative photometry
to achieve the necessary color selection.  

	At fainter magnitudes, it is still possible to partially apply our
method to early-type and late-type stars, if not to G-type stars.  
Of course, as has been mentioned several times, one can exclude 
early-type stars based on color data alone, i.e., without proper motions.
RPM discrimination could then be applied to late-type stars, with a
color threshold that depended on the depth of the survey.  First note that
even for very faint stars $(V<18)$ it is possible to obtain
proper motions with errors $\sigma_\mu\sim 5.5\,\masyr$ for $\delta>-33^\circ$
by matching USNO-A \citep{usnoa1,usnoa2} to 2MASS \citep{faint}, and UCAC2 may
be able to improve somewhat on this value for slightly brighter stars
$(R<16)$.  Suppose a survey star has magnitude $V$ and color $V-I$.  If
the star were a dwarf, then it would have absolute magnitude
$M_V= 3.37(V-I)+2.89$ \citep{reid91}.  
A tranverse velocity of $v_\perp\sim 40\,\kms$
would then be detectable at the $4\,\sigma$ level provided that,
\begin{equation}
V-I > {V -10.8\over 3.37}.
\label{eqn:vmilim}
\end{equation}
Hence, for a survey with limiting magnitude $V=14$
such as the {\it Kepler} mission\footnote{
http://www.kepler.arc.nasa.gov/
}
, it would
still be possible to discriminate between giants and dwarfs for
$V-I>0.95$, roughly the blue edge of the clump.  The great majority
of stars in magnitude-limited samples at these red colors are giants.  Hence,
it is important to have an efficient mechanism for sifting through these to 
find the modest number of dwarfs that are viable targets of a transit search.
(Presently, {\it Kepler} plans to make this selection by high-resolution
spectroscopy of $\sim 10^5$ stars.)

\section{Selection of Larger Stars
\label{sec:bigger}}

\begin{figure}
%\epsscale{0.7}
\plotone{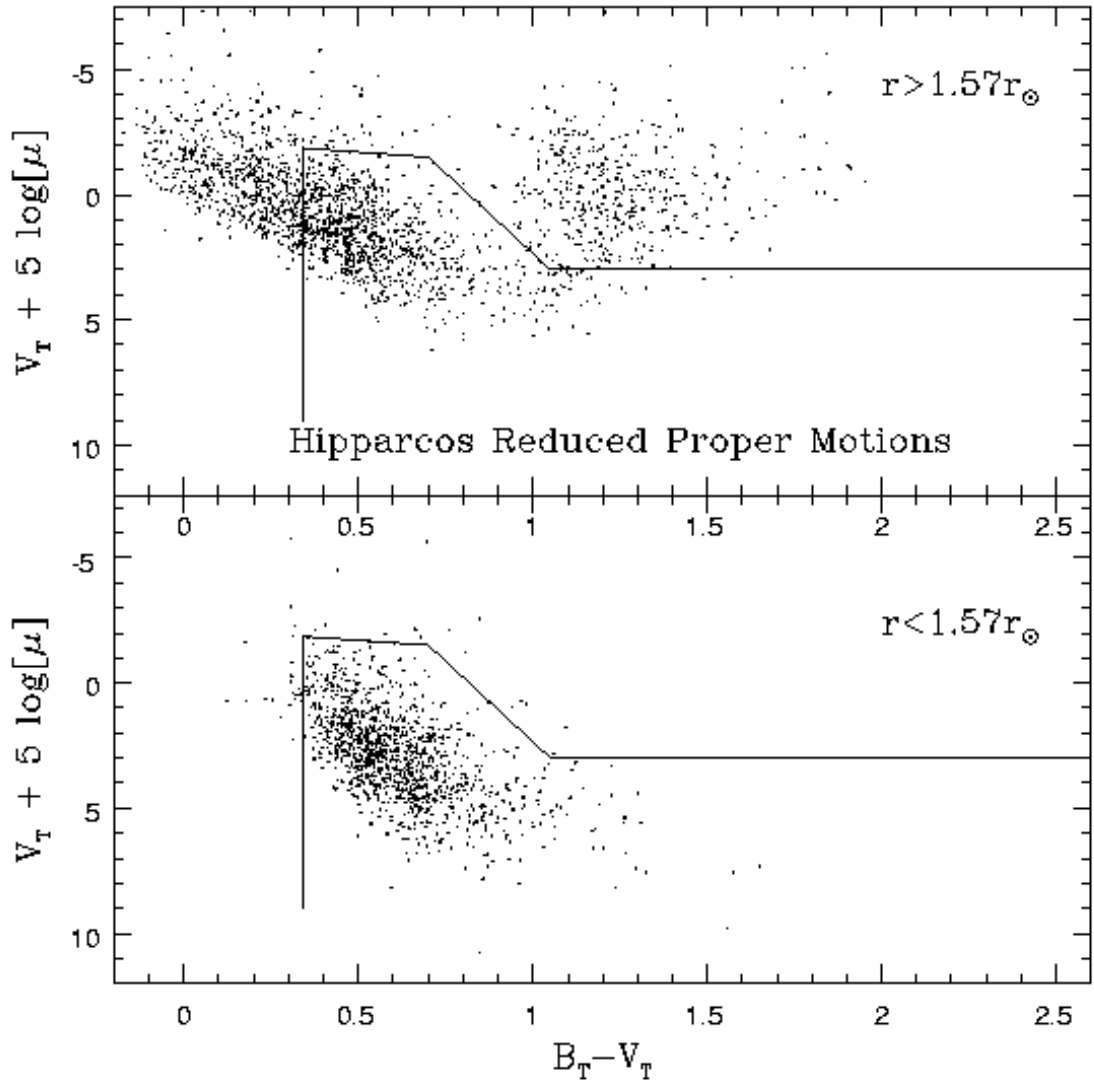}
\caption{\label{fig:hiprpm2}
Hipparcos RPM diagram.  Similar to Fig.~\ref{fig:hiprpm} except that the
division between two panels is now done at $r=1.57\,r_\odot$ (dashed
line in Fig.~\ref{fig:hipcmd}).  The broken line shows our adopted 
criteria to select stars smaller than this radius.  The false rejection
rate is $42/(1157+42)=4\%$.  The contamination rate is probably somewhat
underestimated at $882/(1157+882)=43\%$.
}\end{figure}

	Up to this point, we have implicitly assumed that the transit
survey is systematics-limited, so that there is no point in probing
stars larger than some radius limit, for which we have adopted
$r=1.25\,r_\odot$.  However, we also wish to consider the situation in which 
a survey is photon-limited rather than systematics-limited.  In this case,
if the survey were senstitive to $r=1.25\,r_\odot$ at its magnitude limit,
then it would be sensitive to $r=1.57\,r_\odot$ one magnitude brighter.
In the example of \S~\ref{sec:tycho}, the search among these
bigger stars would be restricted to $V_T<10$.  
The dashed line in Figure~\ref{fig:hipcmd} is the $r=1.57\,r_\odot$
contour.  Figure~\ref{fig:hiprpm2}
shows the derivation of RPM selection criteria for this case.  It is
similar to  Figure~\ref{fig:hiprpm} except that the stars are divided
at $r=1.57\,r_\odot$ rather than $r=1.25\,r_\odot$.  Figure~\ref{fig:tycrpm2}
applies this criterion to the $V_T<10$ sample.  It is similar to
Figure~\ref{fig:tycrpm} except for the brighter magnitude 
limit and larger radius
selection.  The method is slightly less efficient for these bigger stars,
rejecting only 57\% of the stars remaining after a blue color cut has 
removed an initial 22\%.  Recall that the corresponding numbers for
the smaller stars were 60\% and 25\%.
% false rejection (42/1157+42)

\section{Discussion
\label{sec:discuss}}

	The RPM selection function is very efficient at removing early-type
and late-type stars.  For the former, it reduces to a simple color cut
that is similar to the ones already in common use.  For the latter, its
robustness derives from the large gap between the giant branch and the MS.
The method is relatively inefficient at excluding evolved G-type stars
because here the gap is much smaller.  However, this residual G-star 
contamination is modest relative to the contaminants that are efficiently
eliminated.

	Given that the primary new gain from this method is the elimination
of red giant contaminants, one might ask how difficult it would be to
weed these out by other means.  Stellar transits of giants must be very
finely tuned in order to mimic planetary transits.  For example, the eclipse
due to a star in a 0.1 AU (12 day) orbit about a $10\,r_\odot$ giant 
would last $\sim 1\,$day if the inclination were anywhere near $90^\circ$.
Only if it traced a chord within $1\%$ of the giant's limb would the
eclipse be short enough to be confused with a planetary transit.  If the
data were of sufficient quality, the rounded shape of the lightcurve
could reveal the stellar nature of the event.  

	A more difficult problem is caused by triple systems containing
a giant and a MS eclipsing binary.  If the binary separation is bigger
than the radius of the giant, it is extremely difficult to distinguish
this system from a planet transiting a dwarf star using the lightcurve
alone.  Indeed, even spectroscopic observations designed to recognize
the $\sim 10\,\kms$ reflex motion due to an M-star companion would most
likely fail to detect such a contaminant because the giant would not
be markedly changing its RV.  Only a high signal-to-noise ratio spectrum
would succeed in detecting the spectroscopic signature of the companion pair
contributing $\sim 1\%$ of the light.  These difficult contaminants are
easily removed using our method.

	Perhaps the best aspect of our method in this respect is that
it not only removes the need for followup observations of contaminant
events, it removes the need to even analyze the lightcurves of the
great majority of contaminating stars.

	In this paper we have ignored extinction.  The effect of
extinction (if it is not corrected) is to move stars along the reddening
vector in the RPM diagram.  This can essentially only move stars from
being rejected to accepted and not the reverse.  That is, ignoring extinction 
is conservative in that it adds to contamination but does not lead to missing 
legitimate targets.  

\begin{figure}
%\epsscale{0.7}
\plotone{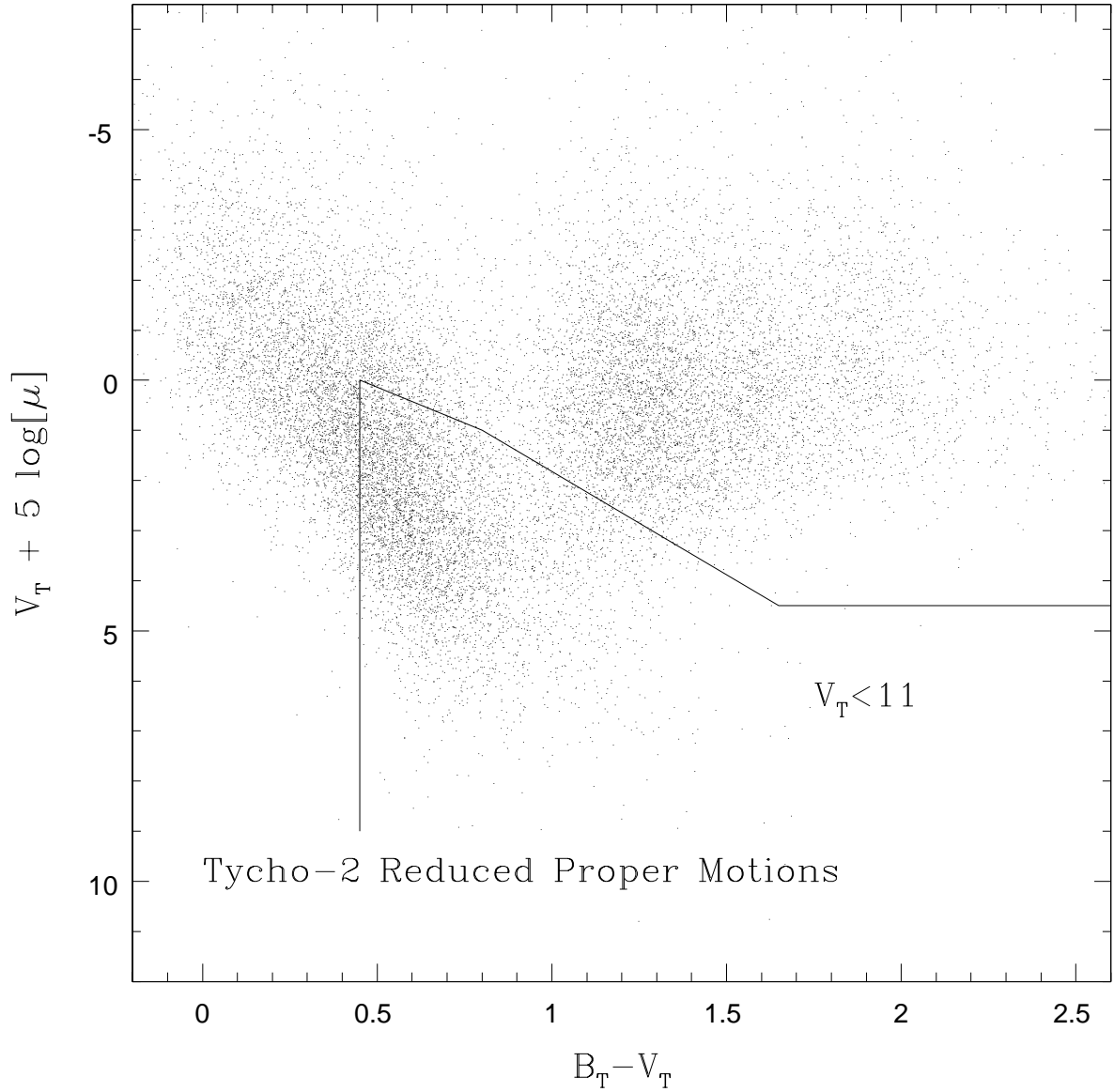}
\caption{\label{fig:tycrpm}
Tycho-2 RPM diagram restricted to stars $V_T<11$.  Only 2\% of stars are
displayed to avoid clutter.  The broken-line selection function (same
as Fig.~\ref{fig:hiprpm}) rejects 70\% of all stars including
25\% early-type (rejected on color grounds alone), 5\% ``G-type''
($0.45<B_T-V_T<0.80$), and 40\% late-type.
}\end{figure}

\begin{figure}
%\epsscale{0.7}
\plotone{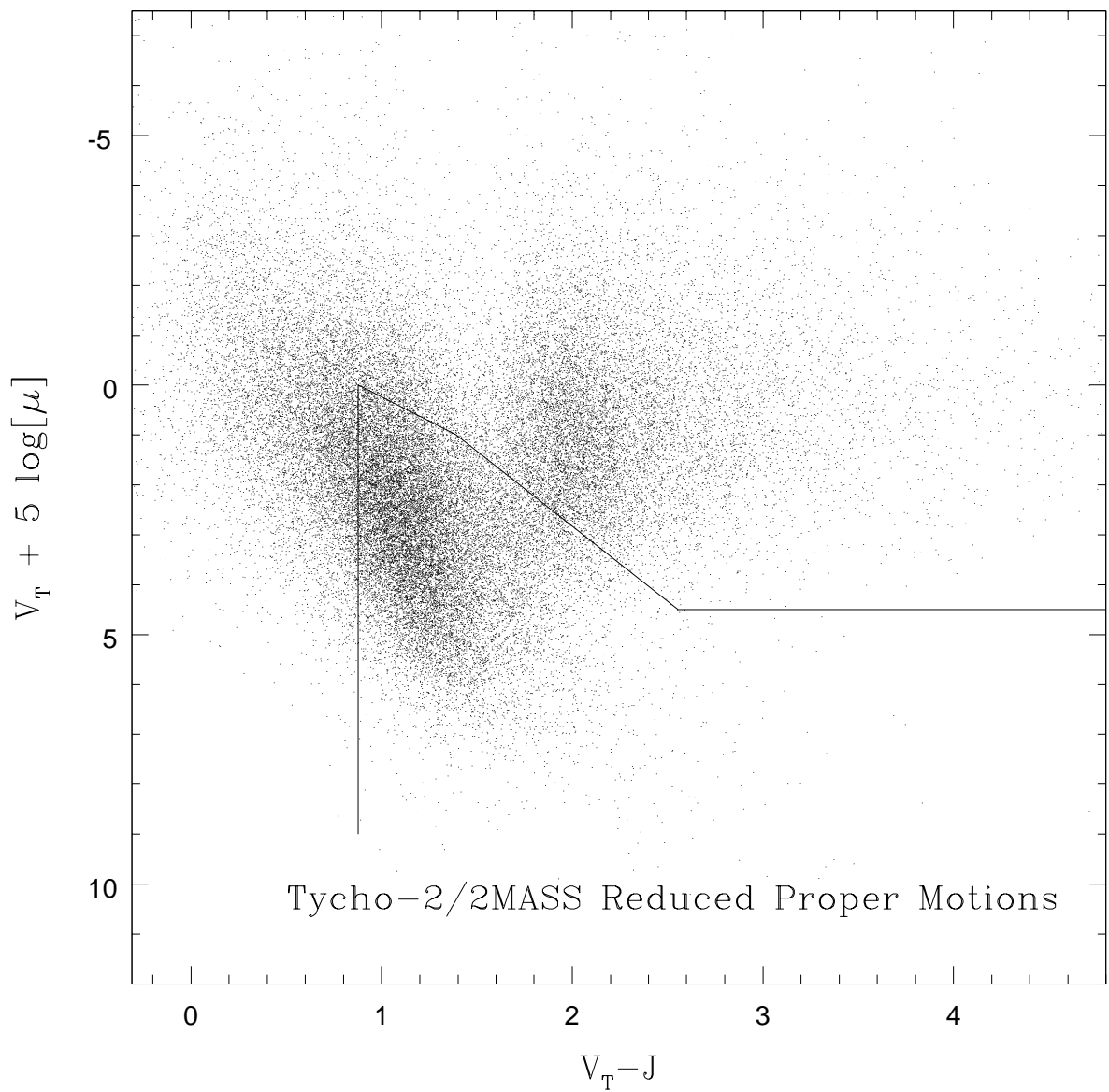}
\caption{\label{fig:2massrpm}
Tycho-2/2MASS RPM diagram.  Similar to Fig.~\ref{fig:tycrpm} except that
by replacing Tycho-2 $B_T$ with 2MASS $J$, we are able to extend coverage
to the Tycho-2 limit, approximately $V\sim 12$.  The broken-line 
selection function is computed by transforming the broken-line in 
Fig.~\ref{fig:tycrpm} using a $(B_T-V_T)/(V_T-J)$ fifth order
color-color relation.
Because the scatter in this relation is only 0.11 mag, the selection is
extremely similar to the Tycho-2-only selection.
}\end{figure}

\begin{figure}
%\epsscale{0.7}
\plotone{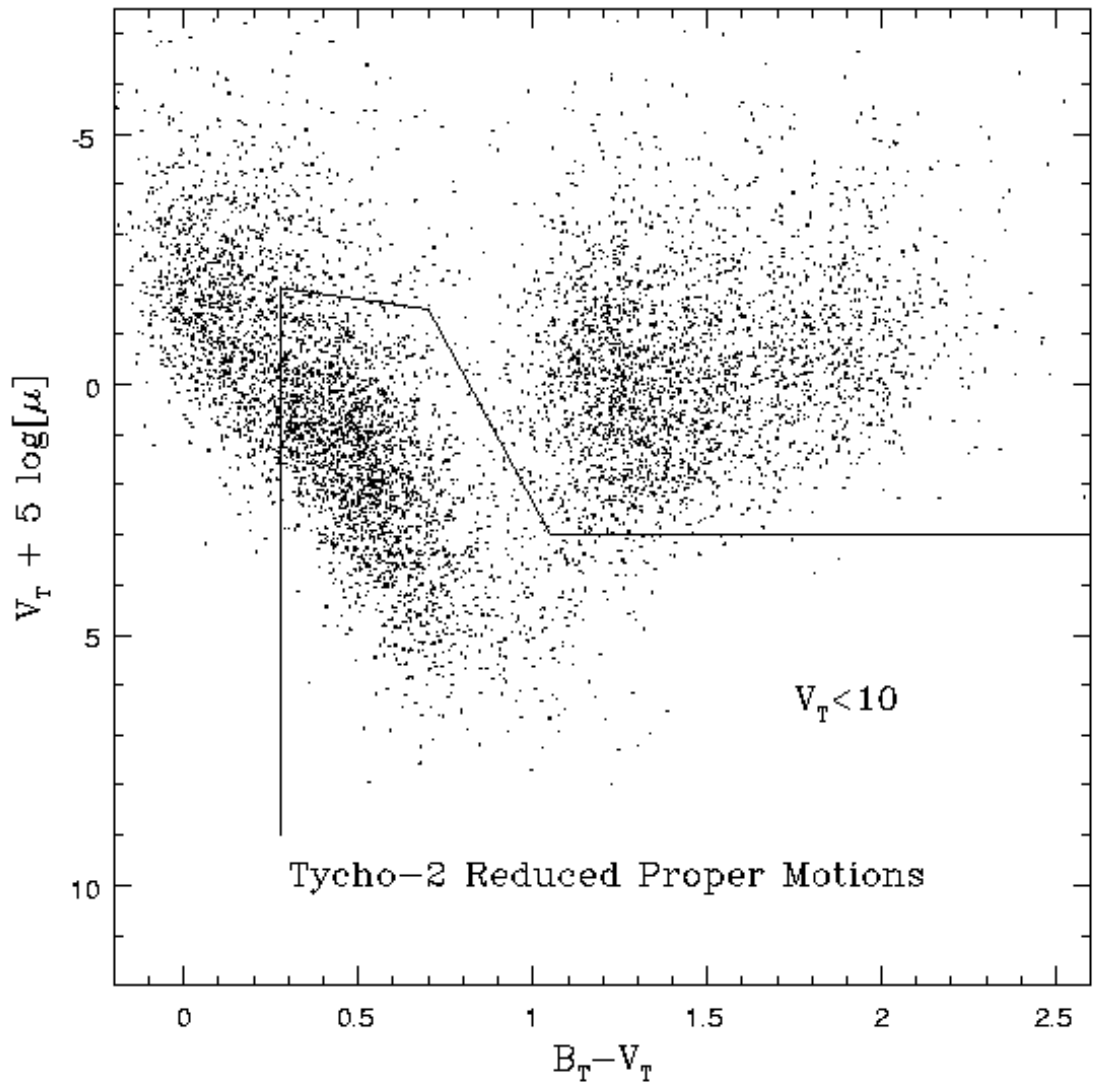}
\caption{\label{fig:tycrpm2}
Tycho-2 RPM diagram.  Similar to Fig.~\ref{fig:tycrpm} except that
first, the sample is restricted to $V_T<10$ and second, the broken line,
which is taken from  Fig.~\ref{fig:hiprpm2}, is designed to select stars
$r<1.57\,r_\odot$.  The efficiency of selection is qualitatively similar
to the case for smaller stars (Fig.~\ref{fig:tycrpm}).
}\end{figure}

	For our primary $V_T<11$ example, shown in Figure \ref{fig:tycrpm},
extinction is actually a very minor effect: target stars $M_{V_T}\ga 4$
always lie $\la 250\,$pc, so that even in the Galactic plane the reddening
is only $E(B_T-V_T)\la 0.06$.  Hence, ignoring extinction has hardly any
effect.  

	As the magnitude limit gets fainter, extinction becomes more
important and so failure to correct for it can lead to significant increases
in contamination.  We caution, however, that accurately correcting for this
effect is not trivial, and that such corrections should therefore be done 
conservatively, i.e., by maintaining
a strong bias against overestimating the extinction.  Even for fields for
which the extinction at infinity is well measured 
(from e.g.\  \citealt{schlegel}), a large fraction of the dust may lie
behind the target stars, which tend to be relatively close.  Thus, the
3-dimensional dust distribution must be modeled.  Moreover, for fields
close to the Galactic plane, i.e., those for which the extinction correction
is most important, the Schegel et al.\ (1998) estimates for extinction at
infinity are less reliable.  Hence, even more care is required in
making the correction.

%\begin{equation}
%\label{eqn:}
%\end{equation}

\acknowledgments This publication makes use of catalogs from the Astronomical 
Data Center at NASA Goddard Space Flight Center, VizieR and SIMBAD databases
operated at CDS, Strasbourg, France, and data products from the Two
Micron All Sky Survey, which is a
joint project of the University of Massachusetts and the IPAC/Caltech,
funded by the NASA and
the NSF.  This work was supported by grant AST 02-01266 from the NSF.

%\clearpage

\clearpage

\begin{deluxetable}{l l r r r r r r r r}
\tabletypesize{\footnotesize}
\tablecaption{Tycho-2 RPM Transit Selection \label{tab:selection}}
\tablewidth{0pt}
\tablehead{

\multicolumn{2}{c}{Sample} &
\colhead{Accept} &
\colhead{Early} &
\colhead{Mid} &
\colhead{Late} &
\colhead{Reject} &
\colhead{Early} &
\colhead{Mid} &
\colhead{Late}\\
\colhead{Mag} &
\colhead{Radius} &
\colhead{$\times 10^3$} &
\colhead{\%} &
\colhead{\%} &
\colhead{\%} &
\colhead{$\times 10^3$} &
\colhead{\%} &
\colhead{\%} &
\colhead{\%} \\
}
\startdata
$V_T<10$ & $r<1.25\,r_\odot$ & 75 & 0 & 79 & 21 & 252 & 40 & 6 & 54 \\
$V_T<11$ & $r<1.25\,r_\odot$ & 257 & 0 & 73 & 27 & 603 & 35 & 8 & 57 \\
All & $r<1.25\,r_\odot$ & 949 & 0 & 68 & 32 & 1231 & 38 & 10 & 52 \\
$V_T<10$ & $r<1.57\,r_\odot$ & 109 & 0 & 77 & 23 & 218 & 33 & 3 & 64 \\

\enddata
\end{deluxetable}


\begin{thebibliography}{}

\bibitem[Brown \& Charbonneau(1999)]{brown99}  Brown, T. M., \&
Charbonneau, D. 1999, \baas, 31, 1534

\bibitem[Charbonneau et al.(2000)]{char00}  Charbonneau, D., Brown, T. M.,
Latham, D. W., \& Mayor, M. 2000, \apj, 529, L45

\bibitem[Charbonneau et al.(2002)]{char02}  Charbonneau, D., Brown, T. M.,
Noyes, R. W., \& Gilliland R. L. 2002, \apj, 568, 377

\bibitem[Cody \& Sasselov(2002)]{cody02}  Cody, A. M., Sasselov, D. D.
2002, \apj, 569, 451

\bibitem[ESA(1997)]{hip} European Space Agency (ESA). 1997, The Hipparcos and
Tycho Catalogues (SP-1200; Noordwijk: ESA)

\bibitem[H{\o}g et al.(2000)]{t2} H{\o}g, E.~et al.\ 2000, \aap, 355, L27.

\bibitem[Howell et al.(2000)]{how00}  Howell, S. B., Everett, M., Davis,
D. R., Weidenschilling, S. J., McGruder, C. H., III, Gelderman, R. 2000,
BAAS, 32, 3203

\bibitem[Hui \& Seager(2002)]{hui02}  Hui, L., Seager, S. 2002, 
\apj, 572, 540

\bibitem[Mallen-Ornelas(2001)]{mal01}  Mallen-Ornelas, G., Seager, S.,
Yee, H. K. C., Minniti, D., Gladders, M. D., Ellison, S., Brown, T.,
Mallen, G. M. 2001, BAAS, 199, 6602


\bibitem[Monet(1996)]{usnoa1} Monet, D.~G.\ 1996, American Astronomical
Society Meeting, 188, 5404.

\bibitem[Monet(1998)]{usnoa2} Monet, D.~G.\ 1998, American Astronomical
Society Meeting, 193, 112003

\bibitem[Pepper, Gould, \& DePoy(2002)]{pgd} Pepper, J., Gould, A., \&
DePoy, D.L.\ 2002, \apj, submitted (astro-ph/0208042)

\bibitem[Reid(1991)]{reid91} Reid, N. 1991, \aj, 102, 1428


\bibitem[Salim \& Gould(2003)]{faint} Salim, S.~\& Gould, A.\ 2003, \apj, 
in press

\bibitem[Schlegel, Finkbeiner, \& Davis(1998)]{schlegel}
Schlegel, D.J., Finkbeiner, D.P., \& Davis, M.\ 1998, \apj, 500, 525

\bibitem[Skrutskie et al.(1997)]{2mass} Skrutskie, M.F.~et al.\ 1997, in The
Impact of Large-Scale Near-IR Sky Survey, ed.\ F.\ Garzon et al.\ (Kluwer:
Dordrecht), p.\ 187

\bibitem[Street et al.(2000)]{str00}  Street, R. A. et al. 2000, in ASP Conf.
Ser., Vol. 219, Euroconference on Disks, Planetesimals and Planets, 
eds. F. Garzon, C. Eiroa, D. de Winter, \& T. J. Mahoney (San Francisco: ASP), 572

\bibitem[Street et al.(2002)]{str02}  Street, R. A. et al. 2002, in ASP Conf.
Ser., Scientific Frontiers in Research on Extrasolar Planets, eds.
D. Deming and S. Seager, in press

\bibitem[Udalski et al.(2002)]{udal02}  Udalski, A. et al. 2002, AcA, 52, 1


\bibitem[van Belle(1999)]{vanb} van Belle, G.T.\ 1999, \pasp, 111, 1515

\bibitem[Zacharias et al.(2000)]{ucac} Zacharias, N.~et al.\ 2000, \aj, 120,
2131
\end{thebibliography}
\end{document}